\documentclass[11pt,draftcls,onecolumn]{IEEEtran}
\usepackage{graphicx,epsfig}  
\usepackage{amsmath}   

\usepackage{amsfonts,amsmath,amssymb}
\usepackage{epic,eepic,eepicemu}
\usepackage{epsf}
\usepackage{graphics}
\usepackage{psfrag}
\usepackage{times}
\usepackage{epsfig}

\usepackage{bbm}
\usepackage{color}

\newcommand{\real}{\mathbb{R}}

\newcommand{\ind}{\mathbb{I}}

\newcommand{\bdm}{\begin{displaymath}}
\newcommand{\edm}{\end{displaymath}}
\newcommand{\bea}{\begin{eqnarray*}}
\newcommand{\eea}{\end{eqnarray*}}
\newcommand{\bean}{\begin{eqnarray}}
\newcommand{\eean}{\end{eqnarray}}

\newcommand{\prob}{\mathbb{P}}
\newcommand{\expec}{\mathbb{E}}
\newcommand{\var}{\mathrm{Var}}

\newcommand{\numnode}{\ensuremath{m}}
\newcommand{\numlevel}{\ensuremath{\ell}}

\newcommand{\broadnum}{\ensuremath{k}}
\newcommand{\nodenum}{\numnode}
\newcommand{\defn}{\ensuremath{: \, = }}
\newcommand{\cdf}{\ensuremath{F}}

\newcommand{\transbit}{\ensuremath{Y}}
\newcommand{\transbitbar}{\ensuremath{\bar{\transbit}}}

\newcommand{\qlevel}{\ensuremath{\alpha^*}}
\newcommand{\qlevmod}{\ensuremath{\widetilde{\alpha}(\chanflip)}}
\newcommand{\thetaopt}{\ensuremath{\theta^*}}
\newcommand{\specpar}{\ensuremath{\beta}}
\newcommand{\const}{\ensuremath{K}}
\newcommand{\myhalf}{\ensuremath{\frac{1}{2}}}
\newcommand{\pdens}{\ensuremath{p_X}}
\newcommand{\normal}{\ensuremath{\mathcal{N}}}
\newcommand{\convdist}{\ensuremath{\stackrel{d}{\rightarrow}}}
\newcommand{\convas}{\ensuremath{\stackrel{a.s.}{\rightarrow}}}
\newcommand{\Exs}{\ensuremath{\mathbb{E}}}
\newcommand{\pdif}[2]{\ensuremath{\frac{\partial #1}{\partial #2}}}

\newcommand{\CondDist}[2]{\ensuremath{\mu_{#1}(#2)}}

\newcommand{\myber}{\ensuremath{\operatorname{Ber}}}

\newcommand{\Gfunc}{\ensuremath{G_\nodenum}}
\newcommand{\sampquant}{\ensuremath{\xi}}
\newcommand{\Qfunc}{\ensuremath{\mathcal{Q}_\numlevel}}
\newcommand{\QfuncO}{\ensuremath{\mathcal{Q}_1}}

\newcommand{\EQfunc}{\ensuremath{G_{\nodenum, \ell}}}

\newtheorem{theorem}{Theorem}
\newtheorem{proposition}{Proposition}

\newtheorem{lemma}{Lemma}

\newcommand{\widgraph}[2]{\includegraphics[keepaspectratio,width=#1]{#2}}

\newcommand{\myparagraph}[1]{{\bf{#1}}}

\makeatletter
\long\def\@makecaption#1#2{
        \vskip 0.8ex
        \setbox\@tempboxa\hbox{\small {\bf #1:} #2}
        \parindent 1.5em  
        \dimen0=\hsize
        \advance\dimen0 by -3em
        \ifdim \wd\@tempboxa >\dimen0
                \hbox to \hsize{
                        \parindent 0em
                        \hfil
                        \parbox{\dimen0}{\def\baselinestretch{0.96}\small
                                {\bf #1.} #2
                                }
                        \hfil}
        \else \hbox to \hsize{\hfil \box\@tempboxa \hfil}
        \fi
        }
\makeatother

\makeatletter
\long\def\barenote#1{
    \insert\footins{\footnotesize
    \interlinepenalty\interfootnotelinepenalty
    \splittopskip\footnotesep
    \splitmaxdepth \dp\strutbox \floatingpenalty \@MM
    \hsize\columnwidth \@parboxrestore
    {\rule{\z@}{\footnotesep}\ignorespaces
      #1\strut}}}
\makeatother

\newcommand{\Prob}{\ensuremath{\mathbb{P}}}

\newcommand{\mybeginproof}{\noindent \emph{Proof: $\;$}}
\newcommand{\myendproof}{\hfill $\blacksquare$}

\newcommand{\plainq}{\ensuremath{\alpha}}
\newcommand{\order}{\ensuremath{\mathcal{O}}}

\newcommand{\numobs}{\ensuremath{n}}

\newcommand{\chanflip}{\ensuremath{\epsilon}}

\long\def\comment#1{}

\title{Universal Quantile Estimation with Feedback in the
Communication-Constrained Setting}

\author{Ram Rajagopal$^1$, Martin J. Wainwright$^{1,2}$\\
$^1$ Department of Electrical Engineering and Computer Science \\
$^2$ Department of Statistics \\ University of California,
Berkeley\\ \texttt{\small $\{$ramr,wainwrig$\}$@eecs.berkeley.edu}
\\ }

\begin{document}

\maketitle

\begin{abstract}
We consider the following problem of decentralized statistical
inference: given i.i.d. samples from an unknown distribution, estimate
an arbitrary quantile subject to limits on the number of bits
exchanged.  We analyze a standard fusion-based architecture, in which
each of $\nodenum$ sensors transmits a single bit to the fusion
center, which in turn is permitted to send some number $k$ bits of
feedback.  Supposing that each of $\nodenum$ sensors receives $n$
observations, the optimal centralized protocol yields mean-squared
error decaying as $\order(1/[n m])$.  We develop and analyze the
performance of various decentralized protocols in comparison to this
centralized gold-standard. First, we describe a decentralized protocol
based on $k = \log(\nodenum)$ bits of feedback that is strongly
consistent, and achieves the same asymptotic MSE as the centralized
optimum.  Second, we describe and analyze a decentralized protocol
based on only a single bit ($k=1$) of feedback.  For step sizes
independent of $m$, it achieves an asymptotic MSE of order
$\order[1/(n \sqrt{m})]$, whereas for step sizes decaying as
$1/\sqrt{m}$, it achieves the same $\order(1/[n m])$ decay in MSE as
the centralized optimum.  Our theoretical results are complemented by
simulations, illustrating the tradeoffs between these different
protocols.
\end{abstract}
{\bf{Keywords:}} Decentralized inference; communication constraints;
distributed estimation; non-parametric estimation; quantiles; sensor
networks; stochastic approximation.\barenote{Portions of this work
were presented at the International Symposium on Information Theory,
Seattle, WA, July 2006.}

\section{Introduction}

Whereas classical statistical inference is performed in a centralized
manner, many modern scientific problems and engineering systems are
inherently \emph{decentralized}: data are distributed, and cannot be
aggregated due to various forms of communication constraints.  An
important example of such a decentralized system is a sensor
network~\cite{Chong03}: a set of spatially-distributed sensors collect
data about the environmental state (e.g., temperature, humidity or
light).  Typically, these networks are based on ad hoc deployments, in
which the individual sensors are low-cost, and must operate under very
severe power constraints (e.g., limited battery life).  In statistical
terms, such communication constraints imply that the individual
sensors cannot transmit the raw data; rather, they must compress or
quantize the data---for instance, by reducing a continuous-valued
observation to a single bit---and can transmit only this compressed
representation back to the fusion center.

By now, there is a rich literature in both information theory and
statistical signal processing on problems of decentralized statistical
inference.  A number of researchers, dating back to the seminal paper
of Tenney and Sandell~\cite{Tenney81}, have studied the problem of
hypothesis testing under communication-constraints; see the survey
papers~\cite{Tsitsiklis93,Veeravalli93,Blum97,Viswanathan97,Chamberland04}
and references therein for overviews of this line of work.  The
hypothesis-testing problem has also been studied in the information
theory community, where the analysis is asymptotic and
Shannon-theoretic in nature~\cite{Amari89,Han89}.  A parallel line of
work deals with problem of decentralized estimation.  Work in signal
processing typically formulates it as a quantizer design problem and
considers finite sample behavior~\cite{Ayanoglu90,Gubner93}; in
contrast, the information-theoretic approach is asymptotic in nature,
based on rate-distortion theory~\cite{Zhang88,Han98}.  In much of the
literature on decentralized statistical inference, it is assumed that
the underlying distributions are known with a specified parametric
form (e.g., Gaussian).  More recent work has addressed non-parametric
and data-driven formulations of these problems, in which the
decision-maker is simply provided samples from the unknown
distribution~\cite{NguWaiJor05,Luo05,Han90}.  For instance, Nguyen et
al.~\cite{NguWaiJor05} established statistical consistency for
non-parametric approaches to decentralized hypothesis testing based on
reproducing kernel Hilbert spaces. Luo~\cite{Luo05} analyzed a
non-parametric formulation of decentralized mean estimation, in which
a fixed but unknown parameter is corrupted by noise with bounded
support but otherwise arbitrary distribution, and shown that
decentralized approaches can achieve error rates that are
order-optimal with respect to the centralized optimum.

This paper addresses a different problem in decentralized
non-parametric inference---namely, that of estimating an arbitrary
quantile of an unknown distribution.  Since there exists no unbiased
estimator based on a single sample, we consider the performance of a
network of $\nodenum$ sensors, each of which collects a total of
$\numobs$ observations in a sequential manner.  Our analysis treats
the standard fusion-based architecture, in which each of the
$\nodenum$ sensors transmits information to the fusion center via a
communication-constrained channel.  More concretely, at each
observation round, each sensor is allowed to transmit a single bit to
the fusion center, which in turn is permitted to send some number $k$
bits of feedback.  For a decentralized protocol with $k =
\log(\nodenum)$ bits of feedback, we prove that the algorithm achieves
the order-optimal rate of the best centralized method (i.e., one with
access to the full collection of raw data).  We also consider a
protocol that permits only a single bit of feedback, and establish
that it achieves the same rate. This single-bit protocol is
advantageous in that, with for a fixed target mean-squared error of
the quantile estimate, it yields longer sensor lifetimes than either
the centralized or full feedback protocols.

The remainder of the paper is organized as follows.  We begin in
Section~\ref{SecDecent} with background on quantile estimation, and
optimal rates in the centralized setting.  We then describe two
algorithms for solving the corresponding decentralized version, based
on $\log(\nodenum)$ and $1$ bit of feedback respectively, and provide
an asymptotic characterization of their performance.  These
theoretical results are complemented with empirical simulations.
Section~\ref{SecTheory} contains the analysis of these two algorithms.
In Section~\ref{SecExt}, we consider various extensions, including the
case of feedback bits $\ell$ varying between the two extremes, and the
effect of noise on the feedforward link.  We conclude in
Section~\ref{SecDiscussion} with a discussion.

\section{Problem Set-up and Decentralized Algorithms}
\label{SecDecent}

In this section, we begin with some background material on
(centralized) quantile estimation, before introducing our
decentralized algorithms, and stating our main theoretical results.

\subsection{Centralized Quantile Estimation}
\label{SecBackground}
We begin with classical background on the problem of quantile
estimation (see Serfling~\cite{Serfling80} for further details).
Given a real-valued random variable $X$, let \mbox{$F(x) \defn \Prob[X
\leq x]$} be its cumulative distribution function (CDF), which is
non-decreasing and right-continuous.  For any $0 < \plainq < 1$, the
$\plainq^{th}$-quantile of $X$ is defined as $F^{-1}(\plainq) \; = \;
\theta(\plainq) \defn \inf \left \{ x \in \real \; \mid \; F(x) \geq
\plainq \right \}$.  Moreover, if $F$ is continuous at $\plainq$, then
we have $\plainq = F(\theta(\plainq))$.  As a particular example, for
$\plainq = 0.5$, the associated quantile is simply the median.

Now suppose that for a fixed level $\qlevel \in (0,1)$, we wish to
estimate the quantile $\thetaopt = \theta(\qlevel)$.  Rather than
impose a particular parameterized form on $F$, we work in a
non-parametric setting, in which we assume only that the distribution
function $F$ is differentiable, so that $X$ has the density function
$p_X(x) = F'(x)$ (w.r.t Lebesgue measure), and moreover that $p_X(x) >
0$ for all $x \in \real$.  In this setting, a standard estimator for
$\thetaopt$ is the \emph{sample quantile} $\sampquant_N(\qlevel) \defn
F_N^{-1}(\qlevel)$ where $F_N$ denotes the empirical distribution
function based on i.i.d. samples $(X_1, \ldots, X_N)$.  Under the
conditions given above, it can be shown~\cite{Serfling80} that
$\sampquant_N(\qlevel)$ is strongly consistent for $\thetaopt$ (i.e.,
$\sampquant_N \convas \thetaopt$), and moreover that asymptotic
normality holds
\begin{eqnarray}
\label{EqnStandAsymp}
\sqrt{N} (\sampquant_N - \thetaopt) & \convdist & \normal \left(0,
\frac{\qlevel(1-\qlevel)}{p^2_X(\thetaopt)}\right),
\end{eqnarray}
so that the asymptotic MSE decreases as $\order(1/N)$, where $N$ is
the total number of samples.  Although this $1/N$ rate is optimal, the
precise form of the asymptotic variance~\eqref{EqnStandAsymp} need not
be in general; see Zielinski~\cite{Zie04} for in-depth discussion of
the optimal asymptotic variances that can be obtained with variants of
this basic estimator under different conditions.

\subsection{Distributed Quantile Estimation}

\begin{figure}
\centering
\psfrag{#x#}{$X \sim  F(\cdot)$}
\psfrag{#m#}{$\nodenum$}
\psfrag{#fus#}{Fusion center}
\widgraph{.32\textwidth}{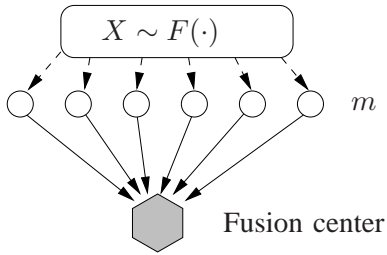}
\caption{Sensor network for quantile estimation with $\nodenum$
sensors.  Each sensor is permitted to transmit a $1$-bit message to
the fusion center; in turn, the fusion center is permitted to
broadcast $k$ bits of feedback.}
\label{FigNetwork}
\end{figure}

We consider the standard network architecture illustrated in
Figure~\ref{FigNetwork}. There are $\nodenum$ sensors, each of which
has a dedicated two-way link to a fusion center.  We assume that each
sensor $i \in \{1, \ldots, \nodenum\}$ collects independent samples
$X(i)$ of the random variable $X \in \real$ with distribution function
\mbox{$\cdf(\theta) \defn \Prob[X \leq \theta]$.}  We consider a
sequential version of the quantile estimation problem, in which sensor
$i$ receives measurements $X_n(i)$ at time steps $n=0,1,2, \ldots$,
and the fusion center forms an estimate $\theta_n$ of the quantile.
The key condition---giving rise to the decentralized nature of the
problem---is that communication between each sensor and the central
processor is constrained, so that the sensor cannot simply relay its
measurement $X(i)$ to the central location, but rather must perform
local computation, and then transmit a summary statistic to the fusion
center. More concretely, we impose the following restrictions on the
protocol.  First, at each time step $n = 0,1,2, \ldots$, each sensor
$i = 1, \ldots, \nodenum$ can transmit a single bit $\transbit_n(i)$
to the fusion center.  Second, the fusion center can broadcast
$\broadnum$ bits back to the sensor nodes at each time step.  We
analyze two distinct protocols, depending on whether \mbox{$\broadnum
= \log(\numnode)$} or \mbox{$\broadnum = 1$.}

\subsection{Protocol specification}

For each protocol, all sensors are initialized with some fixed
$\theta_0$.  The algorithms are specified in terms of a constant
$\const > 0$ and step sizes $\epsilon_n > 0$ that satisfy the
conditions
\begin{equation}
\label{EqnStepSize}
\sum_{n=0}^{\infty}\epsilon_n = \infty  \qquad \mbox{and} \qquad
\sum_{n=0}^{\infty}\epsilon_n^2 < \infty.
\end{equation}
The first condition ensures infinite travel (i.e., that the sequence
$\theta_\numobs$ can reach $\thetaopt$ from any starting condition),
whereas the second condition (which implies that $\epsilon_\numobs
\rightarrow 0$) is required for variance reduction.  A standard choice
satisfying these conditions---and the one that we assume herein---is
$\epsilon_\numobs = 1/\numobs$.  With this set-up, the
$\log(\nodenum)$-bit scheme consists of the steps given in
Table~\ref{TabMbf}.
\newcommand{\framewidth}{0.98\textwidth}
\newcommand{\parboxwidth}{0.96\textwidth}
\begin{table}[h]
\begin{center}
\framebox[\framewidth]{
\parbox{\parboxwidth}{ {\bf{Algorithm: Decentralized quantile
      estimation with $\log(\numnode)$-bit feedback} }

Given  $\const > 0$ and variable step sizes $\epsilon_n > 0$:
\begin{enumerate}
\item[(a)] \emph{Local decision:} each sensor computes the binary
decision
\begin{eqnarray}
\label{EqnBinaryDecFull}
\transbit_{n+1}(i) \equiv \transbit_{n+1}(i; \theta_n) \; \defn \;
\ind(X_{n+1}(i) \leq \theta_n),
\end{eqnarray}
and transmits it to the fusion center.
\item[(b)] \emph{Parameter update:} the fusion center updates its
current estimate $\theta_{n+1}$ of the quantile parameter as
follows:
\begin{eqnarray}
\label{EqnMbf} \theta_{n+1} & = & \theta_n + \epsilon_n \const
\left(\qlevel - \frac{\sum_{i=1}^{\numnode}
\transbit_{n+1}(i)}{\numnode}\right)
\end{eqnarray}
\item[(c)] \emph{Feedback:} the fusion broadcasts the $\numnode$
received bits $\{\transbit_{n+1}(1), \ldots, \transbit_{n+1}(\numnode)
\}$ back to the sensors.  Each sensor can then compute the updated
parameter $\theta_{n+1}$.
\end{enumerate}
}
}
\end{center}
\caption{Description of the $\log(\numnode)$-bf algorithm.}
\label{TabMbf}
\end{table}
%
Although the most straightforward feedback protocol is to broadcast
back the $\numnode$ received bits $\{\transbit_{n+1}(1), \ldots,
\transbit_{n+1}(\numnode) \}$, as described in step (c), in fact it
suffices to transmit only the $\log(\numnode)$ bits required to
perfectly describe the binomial random variable $\sum_{i=1}^{\numnode}
\transbit_{n+1}(i)$ in order to update $\theta_n$.  In either case,
after the feedback step, each sensor knows the value of the sum
$\sum_{i=1}^{\numnode} \transbit_{n+1}(i)$, which (in conjunction with
knowledge of $\numnode$, $\qlevel$ and $\epsilon_n$) allow it to
compute the updated parameter $\theta_{n+1}$.  Finally, knowledge of
$\theta_{n+1}$ allows each sensor to then compute the local
decision~\eqref{EqnBinaryDecFull} in the following round.

%
\begin{table}[h]
\begin{center}
\framebox[\framewidth]{
\parbox{\parboxwidth}{
{\bf{Algorithm: Decentralized quantile estimation with $1$-bit
feedback} }

Given  $\const_\nodenum > 0$ (possibly depending on number of
sensors $\nodenum$) and variable step sizes $\epsilon_n > 0$:
\begin{enumerate}
\item[(a)] \emph{Local decision:} each sensor computes the binary
decision
\begin{eqnarray}
\label{Eqn1bfLocalDec}
\transbit_{n+1}(i) & = & \ind(X_{n+1}(i) \leq \theta_n)
\end{eqnarray}
and transmits it to the fusion center.

\item[(b)] \emph{Aggregate decision and parameter update:} The fusion
center computes the aggregate decision
\begin{eqnarray}
\label{EqnAggDec}
Z_{n+1} & = & \ind \left( \frac{\sum_{i=1}^{\numnode}
\transbit_{n+1}(i)}{\numnode} \leq \qlevel \right),
\end{eqnarray}
and uses it update the parameter according to
\begin{eqnarray}
\label{Eqn1bf}
\theta_{n+1} & = & \theta_n + \epsilon_n \const_m \left(Z_{n+1} -
\specpar \right)
\end{eqnarray}
where the constant $\specpar$ is chosen as
\begin{eqnarray}
\label{EqnDefnSpecPar}
\specpar & = & \sum_{i=0}^{\lfloor \numnode \qlevel \rfloor} {\numnode
 \choose i} (\qlevel)^i \left(1-\qlevel \right)^{\numnode-i}.
\end{eqnarray}

\item[(c)] \emph{Feedback:} The fusion center broadcasts the aggregate
decision $Z_{n+1}$ back to the sensor nodes (one bit of feedback).  Each
sensor can then compute the updated parameter $\theta_{n+1}$.
\end{enumerate}
}
}
\end{center}
\caption{Description of the $1$-bf algorithm.}
\label{Tab1bf}
\end{table}

The 1-bit feedback scheme detailed in Table~\ref{Tab1bf} is similar,
except that it requires broadcasting only a single bit ($Z_{n+1}$),
and involves an extra step size parameter $K_m$, which is specified in
the statement of Theorem~\ref{Thm1bf}. After the feedback step of the
1-bf algorithm, each sensor has knowledge of the aggregate decision
$Z_{n+1}$, which (in conjunction with $\epsilon_n$ and the constant
$\specpar$) allow it to compute the updated parameter $\theta_{n+1}$.
Knowledge of this parameter suffices to compute the local
decision~\eqref{Eqn1bfLocalDec}.

\subsection{Convergence results}

We now state our main results on the convergence behavior of these two
distributed protocols.  In all cases, we assume the step size choice
$\epsilon_n = 1/n$.  Given fixed $\qlevel \in (0,1)$, we use
$\thetaopt$ to denote the $\qlevel$-level quantile (i.e., such that
$\Prob(X \leq \thetaopt) = \qlevel$); note that our assumption of a
strictly positive density guarantees that $\thetaopt$ is unique.
\begin{theorem}[$\numnode$-bit feedback]
\label{ThmMbf}
For any $\qlevel \in (0,1)$, consider a random sequence $\{\theta_n\}$
generated by the $\numnode$-bit feedback protocol.  Then

\noindent (a) For all initial conditions $\theta_0$, the sequence
$\theta_n$ converges almost surely to the $\qlevel$-quantile
$\thetaopt$.

\noindent (b) Moreover, if the constant $\const$ is chosen to
satisfy $\pdens(\thetaopt) \, \const
> \myhalf$, then
\begin{equation}
\label{EqnMbfAsymp}
\sqrt{n} \, (\theta_n-\theta^*) \convdist \normal \left(0,
\frac{\const^2 \; \qlevel \, (1-\qlevel)}{ \big[2 \const
p_X(\thetaopt) - 1\big]} \; \frac{1}{\numnode}\right),
\end{equation}
so that the asymptotic MSE is $O(\frac{1}{\numnode n})$.
\end{theorem}

\noindent \emph{Remarks:} After $n$ steps of this decentralized
protocol, a total of $N = \numobs \numnode$ observations have been
made, so that our discussion in Section~\ref{SecBackground} dictates
(see equation~\eqref{EqnStandAsymp}) that the optimal asymptotic MSE
is $O(\frac{1}{n \numnode})$.  Interestingly, then, the
$\log(\numnode)$-bit feedback decentralized protocol is order-optimal
with respect to the centralized gold standard.

Before stating the analogous result for the 1-bit feedback protocol,
we begin by introducing some useful notation.  First, we define for
any fixed $\theta \in \real$ the random variable
\begin{equation*}
\transbitbar(\theta) \defn \frac{1}{\numnode} \sum_{i=1}^\numnode
\transbit(i; \theta) \; = \; \frac{1}{\numnode} \sum_{i=1}^\numnode
\ind(X(i) \leq \theta).
\end{equation*}
Note that for each fixed $\theta$, the distribution of
$\transbitbar(\theta)$ is binomial with parameters $\numnode$ and
$F(\theta)$.    It is convenient to define the function
\begin{eqnarray}
\label{EqnDefnGfunc}
\Gfunc(r, y) & \defn & \sum_{i=0}^{\lfloor \numnode y \rfloor}
{\numnode \choose i} r^i \; (1-r)^{\numnode -i},
\end{eqnarray}
with domain $(r,y) \in [0,1] \times [0,1]$.  With this notation, we
have
\begin{eqnarray*}
\Prob(\transbitbar(\theta) \leq y) & = & \Gfunc(F(\theta), y).
\end{eqnarray*}
Again, we fix an arbitrary $\qlevel \in (0,1)$ and let $\thetaopt$ be
the associated $\qlevel$-quantile satisfying \mbox{$\Prob(X \leq
\thetaopt) = \qlevel$.}
\begin{theorem}[$1$-bit feedback]
\label{Thm1bf}
Given a random sequence $\{\theta_n\}$ generated by the $1$-bit
feedback protocol, we have
\begin{enumerate}
\item[(a)] For any initial condition, the sequence $\theta_n
\stackrel{a.s.}{\longrightarrow} \thetaopt$.
\item[(b)] Suppose that the step size $\const_\numnode$ is chosen such
 that \mbox{$\const_\numnode > \frac{\sqrt{2 \pi \qlevel (1-\qlevel
 )}}{2 p_X(\thetaopt)\, \sqrt{m}}$,} or equivalently such that
\begin{eqnarray}
\gamma_\nodenum(\thetaopt) & \defn & \const_\numnode \Big
|\pdif{\Gfunc}{r} (r; \qlevel) \big|_{r = \qlevel} \Big | \;
\pdens(\theta^*) \, > \, \frac{1}{2}, \qquad
\end{eqnarray}
then
\begin{equation}
\sqrt{n} \; (\theta_n-\thetaopt) \convdist \normal \left (0,
\frac{\const_\numnode^2 G_\nodenum(\qlevel, \thetaopt)
\big[1-G_\nodenum(\qlevel, \thetaopt)\big]}{2
\gamma_\nodenum(\thetaopt) - 1} \right)
\end{equation}
\item[(c)] If we choose a \emph{constant step size} $\const_\nodenum =
\const$, then as \mbox{$n \rightarrow \infty$,} the asymptotic
variance behaves as
\begin{equation}
\left [\frac{\const^2 \sqrt{2 \pi \qlevel (1-\qlevel )}}{8 \const
p_X(\thetaopt) \sqrt{\nodenum} - 4\sqrt{2 \pi \qlevel (1-\qlevel )}}
\right],
\end{equation}
so that the asymptotic MSE is $O\left(\frac{1}{n
\sqrt{\nodenum}}\right)$.
\item[(d)] If we choose a \emph{decaying step size} $\const_\nodenum =
\frac{\const}{\sqrt{\nodenum}}$, then
\begin{equation}
\label{EqnAsympVarDecay1}
\frac{1}{\nodenum} \: \left [\frac{\const^2 \sqrt{2 \pi \qlevel
(1-\qlevel )}}{8 \const p_X(\thetaopt) - 4\sqrt{2 \pi \qlevel
(1-\qlevel )}} \right],
\end{equation}
 so that the asymptotic MSE is $O\left(\frac{1}{\numobs
 \nodenum}\right)$.
\end{enumerate}
\end{theorem}

\subsection{Comparative Analysis}

It is interesting to compare the performance of each proposed
decentralized algorithm to the centralized performance.  Considering
first the $\log(\numnode)$-bf scheme, suppose that we set $K =
1/p_X(\thetaopt)$.  Using the formula~\eqref{EqnMbfAsymp} from
Theorem~\ref{ThmMbf}, we obtain that the asymptotic variance of the
$\numnode$-bf scheme with this choice of $K$ is given by
$\frac{\qlevel \, (1-\qlevel)}{p^2_X(\thetaopt)} \; \frac{1}{\numnode
\numobs}$, thus matching the asymptotics of the centralized quantile
estimator~\eqref{EqnStandAsymp}.  In fact, it can be shown that the
choice $K = 1/p_X(\thetaopt)$ is optimal in the sense of minimizing
the asymptotic variance for our scheme, when $K$ is constrained by the
stability criterion in Theorem~\ref{ThmMbf}.  In practice, however,
the value $p_X(\thetaopt)$ is typically not known, so that it may not
be possible to implement exactly this scheme. An interesting question
is whether an adaptive scheme could be used to estimate
$p_X(\thetaopt)$ (and hence the optimal $K$ simultaneously), thereby
achieving this optimal asymptotic variance.  We leave this question
open as an interesting direction for future work.

Turning now to the algorithm $1$-bf, if we make the substitution
$\bar{\const} = \const/\sqrt{2\pi\qlevel(1-\qlevel)}$ in
equation~\eqref{EqnAsympVarDecay1}, then we obtain the asymptotic
variance
\begin{equation}
\frac{\pi}{2}\;\frac{\bar{\const}^2 \; \qlevel \, (1-\qlevel)}{ \big[2
\bar{\const} p_X(\thetaopt) - 1\big]} \; \frac{1}{\numnode}.
\end{equation}
Since the stability criterion is the same as that for $m$-bf, the
optimal choice is $\bar{\const} = 1/p_X(\thetaopt)$.  Consequently,
while the $(1/[\numnode \numobs])$ rate is the same as both the
centralized and decentralized $\numnode$-bf protocols, the pre-factor
for the $1$-bf algorithm is $\frac{\pi}{2} \approx 1.57$ times larger
than the optimized $\numnode$-bf scheme.  However, despite this loss
in the pre-factor, the $1$-bf protocol has substantial advantages over
the $\numnode$-bf; in particular, the network lifetime scales as
$O(\numnode)$ compared to $\order(\numnode/\log(\numnode))$ for the
$\log(\numnode)$-bf scheme.

\subsection{Simulation example}

We now provide some simulation results in order to illustrate the two
decentralized protocols, and the agreement between theory and
practice.
\begin{figure*}
\centering
\begin{tabular}{ccc}
\widgraph{0.33\textwidth}{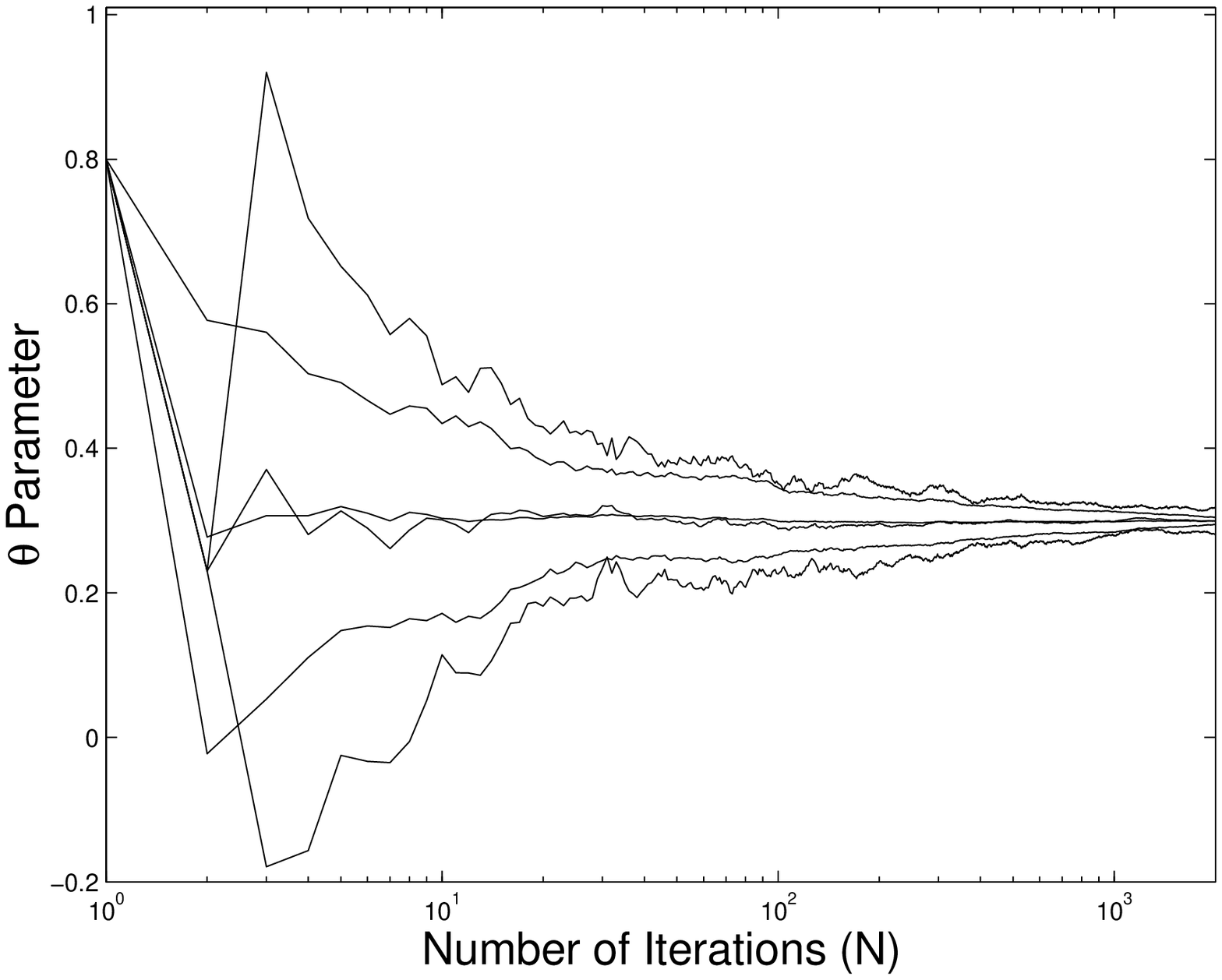} &
\widgraph{0.33\textwidth}{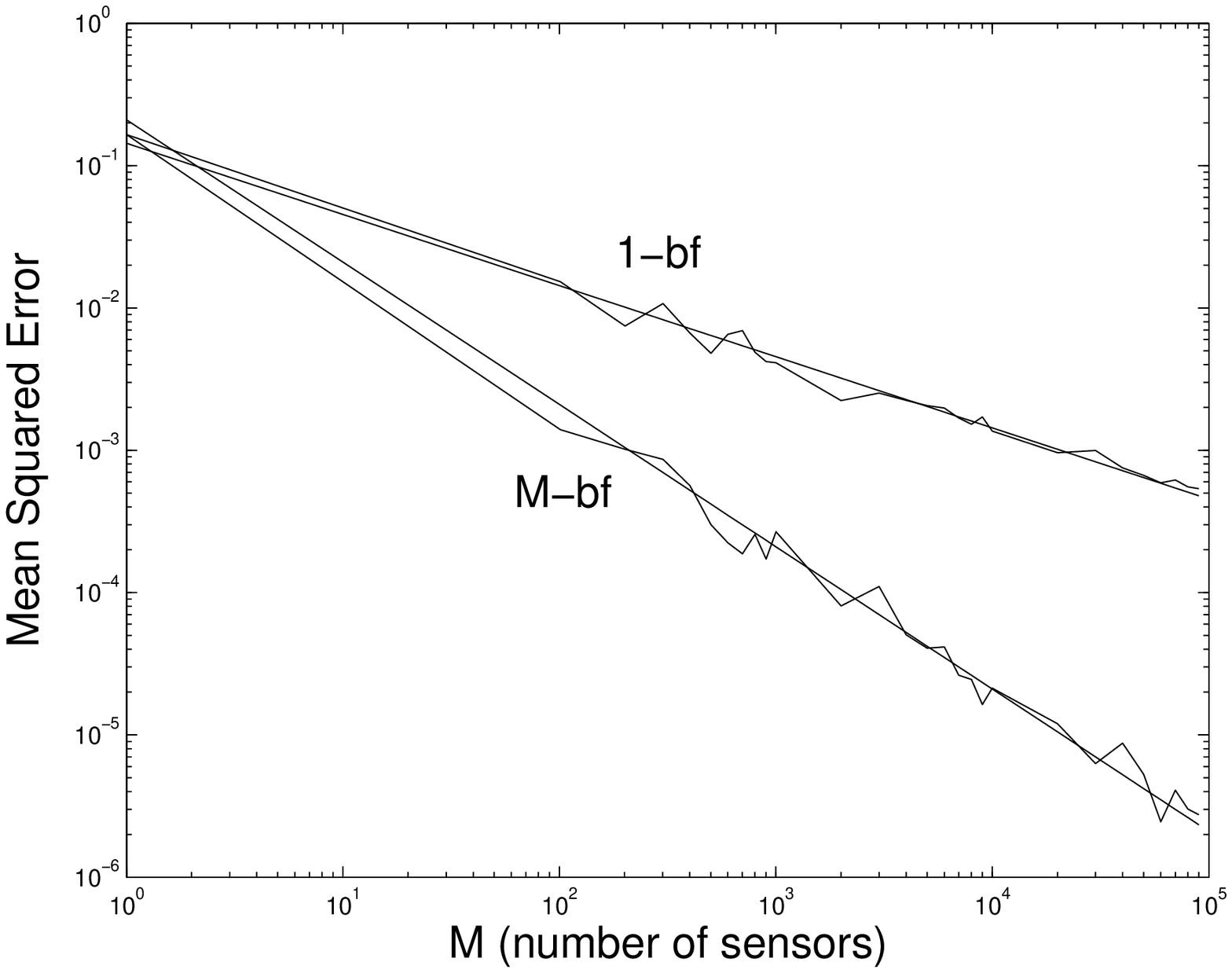} &
\widgraph{0.33\textwidth}{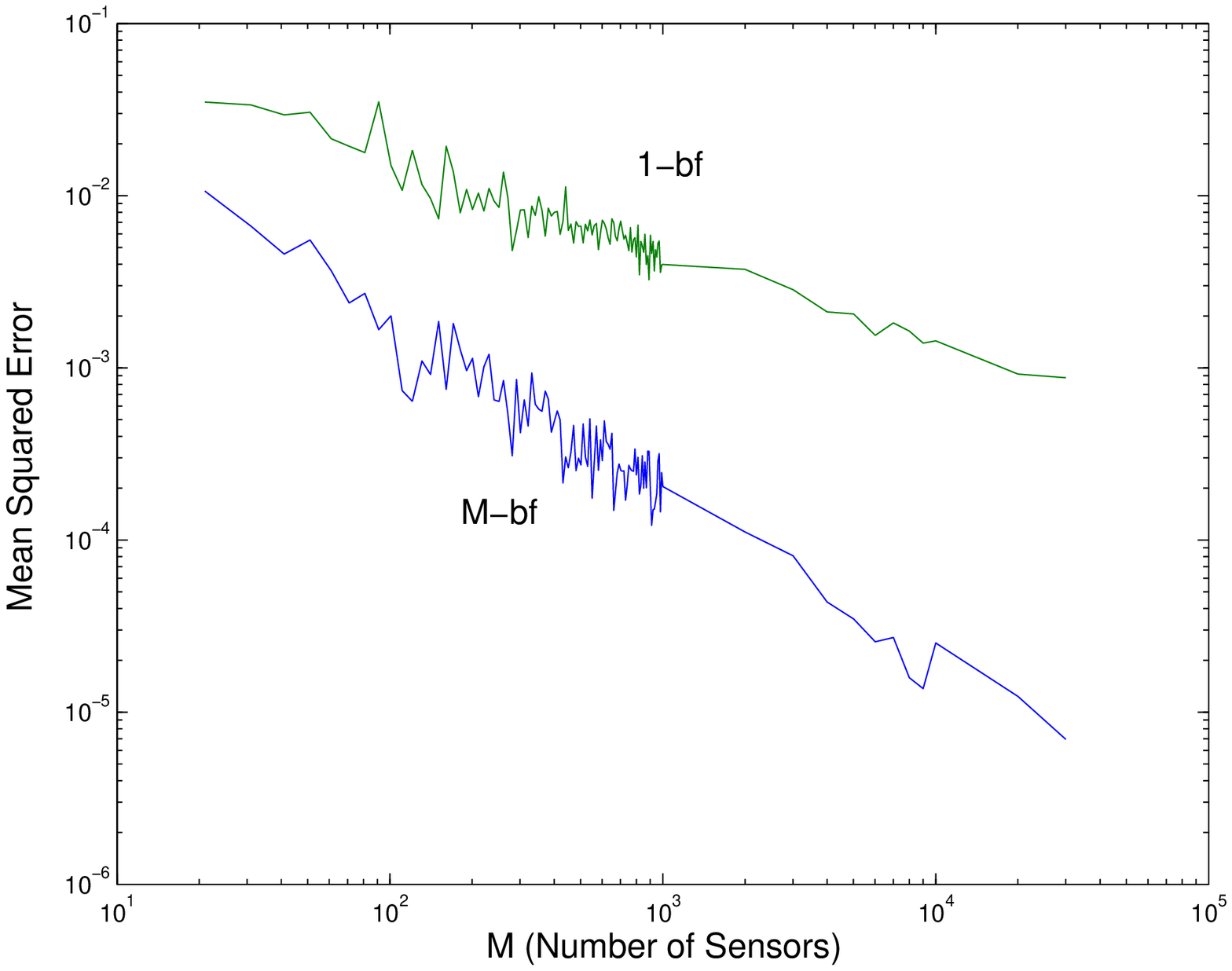} \\
(a) & (b) & (c)
\end{tabular}
\caption{Convergence of $\theta_n$ to $\thetaopt$ with $\numnode=11$
nodes, and quantile level $\qlevel=0.3$.  (b) Log-log plots of the
variance against $\numnode$ for both algorithms ($\log(\numnode)$-bf
and $1$-bf) with constant step sizes, and comparison to the
theoretically-predicted rate (solid straight lines).  \mbox{(c)
Log-log} plots of $\log(\numnode)$-bf with constant step size versus
$1$-bf algorithm with decaying step size.}
\label{FigSims}
\end{figure*}
In particular, we consider the quantile estimation problem when the
underlying distribution (which, of course, is unknown to the
algorithm) is uniform on $[0,1]$.  In this case, we have $p_X(x) = 1$
uniformly for all $x \in [0,1]$, so that taking the constant $\const =
1$ ensures that the stability conditions in both Theorem~\ref{ThmMbf}
and~\ref{Thm1bf} are satisfied.  We simulate the behavior of both
algorithms for $\qlevel = 0.3$ over a range of choices for the network
size $\numnode$.  Figure~\ref{FigSims}(a) illustrates several sample
paths of $\numnode$-bit feedback protocol, showing the convergence to
the correct $\thetaopt$.

For comparison to our theory, we measure the empirical variance by
averaging the error \mbox{$\hat{e}_n = \sqrt{n}(\theta_n -
\thetaopt)$} over $L=20$ runs.  The normalization by $\sqrt{n}$ is
used to isolate the effect of increasing $\numnode$, the number of
nodes in the network. We estimate the variance by running algorithm
for $n=2000$ steps, and computing the empirical variance of
$\hat{e}_n$ for time steps $n = 1800$ through to $n =
2000$. Figure~\ref{FigSims}(b) shows these empirically computed
variances, and a comparison to the theoretical predictions of
Theorems~\ref{ThmMbf} and~\ref{Thm1bf} for constant step size; note
the excellent agreement between theory and practice.  Panel (c) shows
the comparison between the $\log(\numnode)$-bf algorithm, and the
$1$-bf algorithm with decaying $1/\sqrt{\numnode}$ step size.  Here
the asymptotic MSE of both algorithms decays like $1/\numnode$ for
$\log \numnode$ up to roughly $500$; after this point, our fixed
choice of $n$ is insufficient to reveal the asymptotic behavior.

\section{Analysis}

\label{SecTheory}

In this section, we turn to the proofs of Theorem~\ref{ThmMbf}
and~\ref{Thm1bf}, which exploit results from the stochastic
approximation literature~\cite{Kushner97,Benveniste90}.  In
particular, both types of parameter updates~\eqref{EqnMbf}
and~\eqref{Eqn1bf} can be written in the general form
\begin{eqnarray}
\label{EqnGenUpdate}
\theta_{n+1} & = & \theta_n + \epsilon_n H(\theta_n, \transbit_{n+1}),
\end{eqnarray}
where $\transbit_{n+1} = (\transbit_{n+1}(1), \ldots
\transbit_{n+1}(\nodenum))$.  Note that the step size choice
$\epsilon_n = 1/n$ satisfies the conditions in
equation~\eqref{EqnStepSize}. Moreover, the sequence $(\theta_n,
\transbit_{n+1})$ is Markov, since $\theta_n$ and $\transbit_{n+1}$
depend on the past only via $\theta_{n-1}$ and $\transbit_n$.  We
begin by stating some known results from stochastic approximation,
applicable to such Markov sequences, that will be used in our
analysis.

For each fixed $\theta \in \real$, let $\CondDist{\theta}{\, \cdot
\,}$ denote the distribution of $Y$ conditioned on $\theta$.  A key
quantity in the analysis of stochastic approximation algorithms is the
averaged function
\begin{eqnarray}
\label{EqnDefnAveraged}
h(\theta) & \defn & \int H(\theta, y) \CondDist{\theta}{dy} \; = \;
\Exs \left [H(\theta, Y) \mid \theta \right].
\end{eqnarray}
We assume (as is true for our cases) that this expectation exists.
Now the differential equation method dictates that under suitable
conditions, the asymptotic behavior of the update~\eqref{EqnGenUpdate}
is determined essentially by the behavior of the ODE $\frac{d
\theta}{dt} = h(\theta(t))$.

\noindent \myparagraph{Almost sure convergence:} Suppose that the
following \emph{attractiveness condition}
\begin{eqnarray}
\label{EqnAttractive}
h(\theta) \; \left[\theta - \thetaopt \right] & < & 0 \qquad \mbox{for
all $\theta \neq \thetaopt$}
\end{eqnarray}
is satisfied.  If, in addition, the variance \mbox{$R(\theta) \defn
\var[H(\theta; Y) \mid \theta]$} is bounded, then we are are
guaranteed that $\theta_n \convas \thetaopt$ (see \S 5.1 in Benveniste
et al.~\cite{Benveniste90}).

\noindent \myparagraph{Asymptotic normality:} In our updates, the
random variables $\transbit_n$ take the form $\transbit_n = g(X_n,
\theta_n)$ where the $X_n$ are i.i.d. random variables.  Suppose that
the following stability condition is satisfied:
\begin{eqnarray}
\label{EqnStable}
\gamma(\thetaopt) & \defn & -\frac{d h}{d \theta} (\thetaopt) \; > \;
\frac{1}{2}.
\end{eqnarray}
Then we have
\begin{eqnarray}
\label{EqnAsympNormal}
\sqrt{n} \, \left(\theta_n - \thetaopt \right) & \convdist & \normal
\left(0, \frac{R(\thetaopt)}{2 \gamma(\thetaopt) -1)} \right)
\end{eqnarray}
See \S 3.1.2 in Benveniste et al.~\cite{Benveniste90} for further details.

\subsection{Proof of Theorem~\ref{ThmMbf}}

\noindent (a) The $\numnode$-bit feedback algorithm is a special case
of the general update~\eqref{EqnGenUpdate}, with $\epsilon_n =
\frac{1}{n}$ and $H(\theta_n, \transbit_{n+1}) = \const \left
[\alpha^* - \frac{1}{\nodenum} \sum_{i=1}^{\nodenum}
\transbit_{n+1}(i; \theta_n) \right]$.  Computing the averaged
function~\eqref{EqnDefnAveraged}, we have
\begin{eqnarray*}
h(\theta) & = & \const \expec \left[ \qlevel - \frac{1}{\nodenum}
\sum_{i=1}^{\nodenum} \transbit_{n+1}(i) \mid \theta_n \right] \\
& = & \const \left(\qlevel - F(\theta_n) \right),
\end{eqnarray*}
where $F(\theta_n) = \Prob(X \leq \theta_n)$.  We then observe that
$\thetaopt$ satisfies the attractiveness
condition~\eqref{EqnAttractive}, since
\begin{eqnarray*}
\left[\theta - \thetaopt\right] h(\theta_n) & = & \const \,
\left[\theta - \thetaopt\right] \left[\qlevel - F(\theta_n)\right] \;
< \; 0
\end{eqnarray*}
for all $\theta \neq \thetaopt$, by the monotonicity of the cumulative
distribution function.  Finally, we compute the conditional variance
of $H$ as follows:
\begin{eqnarray}
R(\theta_n) & = & \const^2 \var \left[\qlevel -
\frac{\sum_{i=1}^{\nodenum } \transbit_{n+1}(i)}{\nodenum} \mid
\theta_n\right] \nonumber \\
\label{EqnVar}
& = & \frac{\const^2}{\nodenum} F(\theta_n)\left[1-F(\theta_n)\right]
\; \leq \; \frac{\const^2}{4\nodenum},
\end{eqnarray}
using the fact that $H$ is a sum of $\numnode$ Bernoulli variables
that are conditionally i.i.d. (given $\theta_n$).  Thus, we can
conclude that $\theta_n \rightarrow \thetaopt$ almost surely.

\noindent (b) Note that $\gamma(\thetaopt) \; = \; -\frac{d h}{d
\theta} (\thetaopt) = \const p_X(\thetaopt) \, > \, \frac{1}{2}$, so
that the stability condition~\eqref{EqnStable} holds.  Applying the
asymptotic normality result~\eqref{EqnAsympNormal} with the variance
$R(\thetaopt) = \frac{\const^2}{\nodenum} \qlevel (1-\qlevel)$
(computed from equation~\eqref{EqnVar}) yields the claim.

\myendproof


\subsection{Proof of Theorem~\ref{Thm1bf}}

This argument involves additional analysis, due to the aggregate
decision~\eqref{EqnAggDec} taken by the fusion center.  Since the
decision $Z_{n+1}$ is a Bernoulli random variable; we begin by
computing its parameter.  Each transmitted bit $\transbit_{n+1}(i)$ is
$\myber(F(\theta_n))$, where we recall the notation \mbox{$F(\theta)
\defn \Prob(X \leq \theta)$.}  Using the
definition~\eqref{EqnDefnGfunc}, we have the equivalences
\begin{subequations}
\label{EqnEquiv}
\begin{eqnarray}
\label{EqnEquivA}
\Prob(Z_{n+1} = 1) & = & \Gfunc(F(\theta_n), \qlevel) \\
\label{EqnEquivB}
\specpar & = & \Gfunc(\qlevel, \qlevel) \; = \; \Gfunc(F(\thetaopt),
\qlevel). \qquad
\end{eqnarray}
\end{subequations}

We start with the following result:
\begin{lemma} \label{f_mon_dec}
For fixed $x \in [0,1]$, the function $f(r) \defn \Gfunc(r, x)$ is
non-negative, differentiable and monotonically decreasing.
\end{lemma}
\mybeginproof Non-negativity and differentiability are immediate. To
establish monotonicity, note that $f(r) = \Prob(\sum_{i=1}^\numnode
Y_i \leq x \numnode)$, where the $Y_i$ are i.i.d.
$\operatorname{Ber}(r)$ variates.  Consider a second
$\operatorname{Ber}(r')$ sequence $Y'_i$ with $r' > r$.  Then the sum
$\sum_{i=1}^\nodenum Y'_i$ stochastically dominates
$\sum_{i=1}^\nodenum Y_i$, so that $f(r) < f(r')$ as required.

\myendproof

To establish almost sure convergence, we use a similar approach as in
the previous theorem.  Using the equivalences~\eqref{EqnEquiv}, we
compute the function $h$ as follows
\begin{eqnarray*}
h(\theta) & = & \const_m \expec\left[Z_{n+1} - \specpar \mid
\theta\right] \\
& = & \const_m \left[\Gfunc(F(\theta), \qlevel) -
\Gfunc(F(\thetaopt), \qlevel) \right].
\end{eqnarray*}
Next we establish the attractiveness condition~\eqref{EqnAttractive}.
In particular, for any $\theta$ such that $F(\theta) \neq
F(\thetaopt)$, we calculate that $h(\theta) \, \left[\theta -
\thetaopt\right]$ is given by
\begin{eqnarray*}
\const_m \Big \{ \Gfunc(F(\theta_n), \qlevel) - \Gfunc(F(\thetaopt),
\qlevel) \Big \} \; \left[\theta_n - \thetaopt\right] & < & 0,
\end{eqnarray*}
where the inequality follows from the fact that $\Gfunc(r, x)$ is
monotonically decreasing in $r$ for each fixed $x \in [0,1]$ (using
Lemma~\ref{f_mon_dec}), and that the function $F$ is monotonically
increasing.  Finally, computing the variance $R(\theta) \defn
\var\left[H(\theta, Y) \mid \theta\right]$, we have
\begin{eqnarray*}
R(\theta) &=& \const_m^2 \Gfunc(F(\theta), \qlevel) \; \left[1 -
\Gfunc(F(\theta) , \qlevel) \right] \; \leq \;
\frac{\const_m^2}{4}
\end{eqnarray*}
since (conditioned on $\theta$), the decision $Z_{n+1}$ is Bernoulli
with parameter $\Gfunc(F(\theta); \qlevel)$.  Thus, we can conclude
that $\theta_n \rightarrow \theta^*$ almost surely.

\noindent (b) To show asymptotic normality, we need to verify the
stability condition.  By chain rule, we have $\frac{
h}{d\theta}(\thetaopt) = \const_m \frac{\partial \Gfunc}{\partial
r}(r, \qlevel) \Big|_{r = F(\theta)} \, p_X(\theta)$.  From
Lemma~\ref{f_mon_dec}, we have $ \frac{\partial \Gfunc}{\partial
r}(F(\theta), \qlevel) < 0$, so that the stability condition holds
as long as $\gamma_\nodenum(\thetaopt) > \frac{1}{2}$ (where
$\gamma_\nodenum$ is defined in the statement).  Thus, asymptotic
normality holds.

In order to compute the asymptotic variance, we need to investigate
the behavior of $R(\thetaopt)$ and $\gamma(\thetaopt)$ as $\nodenum
\rightarrow +\infty$.  First examining $R(\thetaopt)$, the central
limit theorem guarantees that \mbox{$\Gfunc(F(\thetaopt), y)
\rightarrow \Phi\left(\sqrt{\nodenum}
\frac{y-\alpha^*}{\alpha^*(1-\alpha^*)}\right)$.}  Consequently, we
have
\begin{equation*}
R(\thetaopt) \; = \; \const_m^{2} \Gfunc(F(\thetaopt), \qlevel) \,
\left[1 - \Gfunc(F(\thetaopt), \qlevel)\right] \rightarrow
\frac{\const_m^{2}}{4}.
\end{equation*}

We now turn to the behavior of $\gamma(\thetaopt)$. We first prove a
lemma to characterize the asymptotic behavior of $\Gfunc (r,
\qlevel)$:
\begin{lemma}\label{f_mon_rate}
\noindent (a) The partial derivative of $\Gfunc (r, x)$ with respect
to $r$ is given by:
\begin{eqnarray}
\label{EqnGpartial}
\frac{\partial \Gfunc (r, x)}{\partial r}
& = & \frac{\Exs[X \ind(X \leq x \numnode)] - \Exs[X] \Exs[\ind(X
\leq x \numnode)]}{r(1-r)},
\end{eqnarray}
where $X$ is binomial with parameters $(\numnode, x)$, and mean
$\Exs[X] = x \numnode$.

\noindent (b)  Moreover, as $\nodenum \rightarrow +\infty$, we have
\begin{eqnarray*}
\frac{\partial \Gfunc (r, \qlevel)}{\partial r} \big|_{r =
F(\thetaopt)} & \rightarrow & -\sqrt{\frac{\nodenum}{2\pi \qlevel
(1-\qlevel)}}.
\end{eqnarray*}
\end{lemma}
\mybeginproof (a) Computing the partial derivative, we have
\begin{eqnarray*}
\frac{\partial \Gfunc (r, x)}{\partial r} & =& \sum_{i=0}^{\lfloor
\nodenum \qlevel \rfloor} {\nodenum \choose i} \left [ i r^{i-1}
(1-r)^{\nodenum-i} - (\nodenum-i) r^{i} (1-r)^{\nodenum-i-1} \right]\\
& = &\frac{1}{r(1-r)} \sum_{i=0}^{\lfloor \nodenum x \rfloor} \left(
\begin{array}{c} \nodenum \\ i
 \end{array}\right)(i- \nodenum r)r^{i}(1-r)^{\nodenum-i} \\
       & = & \frac{1}{r(1-r)} \left(  \sum_{i=0}^{\lfloor \nodenum x \rfloor}
 \left( \begin{array}{c} \nodenum \\ i
 \end{array}\right)r^{i}(1-r)^{\nodenum-i}-\nodenum r\sum_{i=0}^{\lfloor  \nodenum x \rfloor}
 \left( \begin{array}{c} \nodenum \\ i
 \end{array}\right)r^{i}(1-r)^{\nodenum -i}\right)\\
      & = &  \frac{1}{r(1-r)} \left(\expec[X \ind(X\leq
       \nodenum x)]-\expec[X] \expec[\ind(X\leq  \nodenum x)]\right) \label{d_fr},
\end{eqnarray*}
as claimed.

\newcommand{\specn}{\ensuremath{\sqrt{\numnode}}}

\noindent (b) We derive this limiting behavior by applying classical
asymptotics to the form of $\frac{\partial \Gfunc (r,
\qlevel)}{\partial r}$ given in part (a).  Defining $Z_\numnode =
\frac{X - \qlevel \numnode}{\sqrt{\numnode}}$, the central limit
theorem yields that:
\begin{eqnarray}
\label{EqnNormAs}
Z_\numnode &\convdist& Z \sim N(0, a) \\
a & \defn & \qlevel\,(1-\qlevel) \nonumber
\end{eqnarray}
Moreover, in this binomial case, we actually have $\Exs[|Z_\numnode|]
\rightarrow \Exs[|Z|] = \sqrt{\frac{2 a}{\pi}}$.

First, since $\Exs[X] = \qlevel \numnode$ and $\Exs[\ind(X \leq
\qlevel \numnode)] \rightarrow \frac{1}{2}$ by the CLT, we have
\begin{eqnarray}
\label{EqnLimitOne}
\Exs[X] \; \Exs[\ind(X \leq \qlevel \numnode)] & \rightarrow &
\frac{\qlevel \numnode}{2}.
\end{eqnarray}
Let us now re-write the first term in the
representation~\eqref{EqnGpartial} of $\frac{\partial \Gfunc (r,
\qlevel)}{\partial r}$ as
\begin{eqnarray}
\expec[X \ind(X\leq \qlevel \numnode )] & = & \qlevel \numnode
\expec[\ind(X \leq \qlevel \numnode)] + \specn \expec[ Z_\numnode \,
\ind(Z_\numnode \leq 0)] \nonumber \\
\label{EqnLimitTwo} & \rightarrow & \frac{\qlevel \numnode}{2} -
\specn \sqrt{\frac{a}{2 \pi}}
\end{eqnarray}
since $\expec[\ind(X \leq \qlevel \numnode)] \rightarrow 1/2$ and
\[
\expec[ Z_\numnode \, \ind(Z_\numnode \leq 0)] \; \rightarrow \;
\Exs[Z \ind(Z \leq 0)] = \frac{1}{2} \Exs[|Z|] = \sqrt{\frac{a}{2
\pi}}.
\]
Putting together the limits~\eqref{EqnLimitOne}
and~\eqref{EqnLimitTwo}, we conclude that $\frac{\partial \Gfunc (r,
\qlevel)}{\partial r} \big|_{r = \qlevel}$ converges to
\begin{eqnarray*}
\frac{1}{\qlevel(1-\qlevel)}\left[ \left \{\frac{\qlevel
\numnode}{2} - \specn \sqrt{\frac{\qlevel \,(1-\qlevel)}{2 \pi}}
\right \}- \frac{\qlevel \numnode}{2}\right] & = &
-\sqrt{\frac{\numnode}{2\pi\qlevel(1-\qlevel)}},
\end{eqnarray*}
as claimed.
\myendproof

Returning now to the proof of the theorem, we use
Lemma~\ref{f_mon_rate} and put the pieces together to obtain that
$\frac{R(\thetaopt)}{2 \const_\numnode \left |\frac{\partial \Gfunc(r,
\thetaopt)}{\partial r} \big |_{r = \qlevel} \right| p_X(\thetaopt) -
1}$ converges to
\begin{eqnarray*}
\frac{\const_\nodenum^{2}/4}{\frac{2 \const_\numnode \sqrt{\numnode}
  p_X(\thetaopt)}{\sqrt{2\pi \qlevel (1-\qlevel)}} - 1} & = &
  \frac{1}{\nodenum} \left [\frac{\const^2 \sqrt{2 \pi \qlevel
  (1-\qlevel )}}{8 K p_X(\thetaopt)-4\sqrt{2 \pi \qlevel (1-\qlevel
  )}} \right],
\end{eqnarray*}
with $K > \frac{\sqrt{2 \pi \qlevel (1-\qlevel )}}{2p_X(\thetaopt)}$
for stability, thus completing the proof of the theorem.

\myendproof

\section{Some extensions}
\label{SecExt}

In this section, we consider some extensions of the algorithms and
analysis from the preceding sections, including variations in the
number of feedback bits, and the effects of noise.

\subsection{Different levels of feedback}

We first consider the generalization of the preceding analysis to the
case when the fusion center some number of bits between $1$ and
$\numnode$.  The basic idea is to apply a quantizer with $2 \numlevel$
levels, corresponding to $\log_2 (2\numlevel)$ bits, on the update of
the stochastic gradient algorithm.  Note that the extremes $\numlevel
= 1$ and $\numlevel = 2^{\numnode-1}$ correspond to the previously
studied protocols.  Given $2 \numlevel$ levels, we partition the real
line as
\begin{equation}
-\infty \; = \; s_{-\numlevel} \; < \; s_{-\numlevel + 1} \; < \;
 \ldots \; < \; s_{\numlevel-1} \; < \; s_{\numlevel} = +\infty,
\end{equation}
where the remaining breakpoints $\{s_{k}\}$ are to be specified.  With
this partition fixed, we define a quantization function $\Qfunc$
\begin{eqnarray}
\Qfunc(X) & \defn & r_k \qquad \mbox{if $X \in (s_k, s_{k+1}]$ for $k
    = -\numlevel, \ldots, \numlevel-1$},
\end{eqnarray}
where the $2 \numlevel$ quantized values $(r_{-\numlevel}, \ldots,
r_{\numlevel-1})$ are to be chosen.  In the setting of the algorithm
to be proposed, the quantizer is applied to binomial random variables
$X$ with parameters $(\numnode, r)$. Recall the function $\Gfunc(r,
x)$, as defined in equation~\eqref{EqnDefnGfunc}, corresponding to the
probability $\Prob[X \leq \numnode x]$.  Let us define a new function
$\EQfunc$, corresponding to the expected value of the quantizer when
applied to such a binomial variate, as follows
\begin{eqnarray}
\label{EqnEqfuncDef}
\EQfunc(r,x) & \defn & \sum_{k = -\numlevel}^{\numlevel-1} r_k \left
\{\Gfunc(r, x-s_{k})-\Gfunc(r, x-s_{k+1})\right \}.
\end{eqnarray}
With these definitions, the general $\log_2(2 \numlevel)$ feedback
algorithm takes the form shown in Table~\ref{TabGen}.
\begin{table}[h]
\begin{center}
\framebox[\framewidth]{
\parbox{\parboxwidth}{ {\bf{Algorithm: Decentralized quantile
estimation with $\log_2(2\numlevel)$-bits feedback} }

Given  $\const_\nodenum > 0$ (possibly depending on number of
sensors $\nodenum$) and variable step sizes $\epsilon_n > 0$:
\begin{enumerate}
\item[(a)] \emph{Local decision:} each sensor computes the binary
decision
\begin{eqnarray}
\label{EqnLocalDecGen}
\transbit_{n+1}(i) & = & \ind(X_{n+1}(i) \leq \theta_n)
\end{eqnarray}
and transmits it to the fusion center.

\item[(b)] \emph{Aggregate decision and parameter update:} The fusion
center computes the quantized aggregate decision variable
\begin{eqnarray}
\label{EqnAggDecGen} Z_{n+1} & = & \Qfunc \left[\qlevel -
\frac{\sum_{i=1}^{\numnode} \transbit_{n+1}(i)}{\numnode} \right],
\end{eqnarray}
and uses it update the parameter according to
\begin{eqnarray}
\label{EqnUpdateGen} \theta_{n+1} & = & \theta_n + \epsilon_n \const_m
\left(Z_{n+1} - \specpar \right)
\end{eqnarray}
where the constant $\specpar$ is chosen as
\begin{eqnarray}
\label{EqnDefnSpecParGen}
\specpar & \defn & \EQfunc(F(\thetaopt), \qlevel).
\end{eqnarray}
\item[(c)] \emph{Feedback:} The fusion center broadcasts the aggregate
quantized decision $Z_{n+1}$ back to the sensor nodes, using its
$\log_2(2 \numlevel)$ bits of feedback.  The sensor nodes can then
compute the updated parameter $\theta_{n+1}$.
\end{enumerate}
}
}
\end{center}
\caption{Description of the general algorithm, with $\log_2(2 \numlevel)$ bits
of feedback.}
\label{TabGen}
\end{table}

In order to understand the choice of the offset parameter $\specpar$
defined in equation~\eqref{EqnDefnSpecParGen}, we compute the expected
value of the quantizer function, when $\theta_n = \thetaopt$, as
follows
\begin{eqnarray*}
\expec\Big[\Qfunc \left[ \qlevel-\frac{ \sum_{i=1}^{\numnode}
\transbit_{n+1}(i)}{\numnode} \right] \; \mid \; \theta_n =
\thetaopt \Big ] &=& \sum_{k = -\numlevel}^{\numlevel-1} r_k \Prob
\left[(\qlevel-s_{k+1}) < \frac{\transbitbar(\thetaopt)}{\numnode}
\leq
(\qlevel-s_{k}) \right] \\
&=& \sum_{k = -\numlevel}^{\numlevel-1} r_k \left
[\Gfunc(F(\thetaopt), \qlevel-s_{k})-\Gfunc(F(\thetaopt),
\qlevel-s_{k+1}) \right] \\
& = & \EQfunc(F(\thetaopt), \qlevel).
\end{eqnarray*}
The following result, analogous to Theorem~\ref{Thm1bf}, characterizes
the behavior of this general protocol:
\begin{theorem}[General feedback scheme]
\label{ThmGen}
Given a random sequence $\{\theta_n\}$ generated by the general
$\log_2(2 \numlevel)$-bit feedback protocol, there exist choices of
partition $\{s_k\}$ and quantization levels $\{r_k \}$ such that:
\begin{enumerate}
\item[(a)] For any initial condition, the sequence $\theta_n
\stackrel{a.s.}{\longrightarrow} \thetaopt$.
\item[(b)] There exists a choice of \emph{decaying step size} (i.e.,
$\const_\nodenum \asymp \frac{1}{\sqrt{\nodenum}}$) such that the
asymptotic variance of the protocol is given by
$\frac{\kappa(\qlevel,\Qfunc)}{\numnode \numobs}$, where the constant
has the form
\begin{eqnarray}
\label{EqnDefnKappa}
\kappa(\qlevel,\Qfunc) & \defn & 2\pi \; \frac{\sum_{k =
    -\numlevel}^{\numlevel-1}r_k^2 \Delta\Gfunc(s_{k},s_{k+1})
  -\specpar^2}{\left(\sum_{k
    =-\numlevel}^{\numlevel-1}r_k\Delta_{\numnode}(s_{k},s_{k+1})\right)^2},
\end{eqnarray}
with
\begin{subequations}
\begin{eqnarray}
\Delta\Gfunc(s_{k},s_{k+1}) &=& \Gfunc(F(\thetaopt),
\qlevel-s_{k})-\Gfunc(F(\thetaopt), \qlevel-s_{k+1}), \quad \mbox{and} \\
\Delta_{\numnode}(s_{k},s_{k+1}) &=& \exp\left(-\frac{\numnode
s_{k}^2}{2\qlevel(1-\qlevel)}\right)-\exp\left(-\frac{\numnode
s_{k+1}^2}{2\qlevel(1-\qlevel)}\right).
\end{eqnarray}
\end{subequations}
\end{enumerate}
\end{theorem}
We provide a formal proof of Theorem~\ref{ThmGen} in the Appendix.
Figure~\ref{FigLevels}(a) illustrates how the constant factor
$\kappa$, as defined in equation~\eqref{EqnDefnKappa} decreases as the
number of levels $\numlevel$ in an uniform quantizer is increased.

In order to provide comparison with results from the previous section,
let us see how the two extreme cases ($1$ bit and $\numnode$ feedback)
can be obtained as special case.  For the $1$-bit case, the quantizer
has $\numlevel = 1$ levels with breakpoints $s_{-1} = -\infty$, $s_0 =
0$, $s_{1} = +\infty$, and quantizer outputs $r_{-1} = 0$ and $r_{1} =
1$. By making the appropriate substitutions, we obtain:
\begin{eqnarray*}
\kappa(\qlevel,\QfuncO) \; = \;  2\pi \; \frac{\Delta\Gfunc(s_{0},s_{1})
  -\specpar^2}{\Delta_{\numnode}(s_{0},s_{1})},  & \qquad \qquad & \specpar^2
\; = \; \EQfunc(F(\thetaopt), \qlevel)^2, \\
\Delta\Gfunc(s_{0},s_{1}) \; = \; \EQfunc(F(\thetaopt), \qlevel) &
\mbox{and} & \Delta_{\numnode}(s_{0},s_{1})) \; = \; 1.
\end{eqnarray*}
By applying the central limit theorem, we conclude that
\[
\Delta\Gfunc(s_{0},s_{1}) -\specpar^2 =\EQfunc(F(\thetaopt),
\qlevel)(1-\EQfunc(F(\thetaopt), \qlevel))\rightarrow 1/4,
\]
as established earlier.  Thus $\kappa(\qlevel,\QfuncO) \rightarrow
\pi/2$ as $\nodenum \rightarrow \infty$, recovering the result of
Theorem~\ref{Thm1bf}.  Similarly, the results for $\nodenum$-bf can be
recovered by setting the parameters
\begin{eqnarray}
r_{k-\numlevel} & = & \qlevel - \frac{\nodenum-k}{\nodenum}, \quad
\mbox{for}
\quad k = 0,...,\nodenum, \qquad \mbox{and} \nonumber \\
s_i &=& r_i.
\end{eqnarray}

\begin{figure}[h]
\begin{center}
\begin{tabular}{cc}
\widgraph{0.5\textwidth}{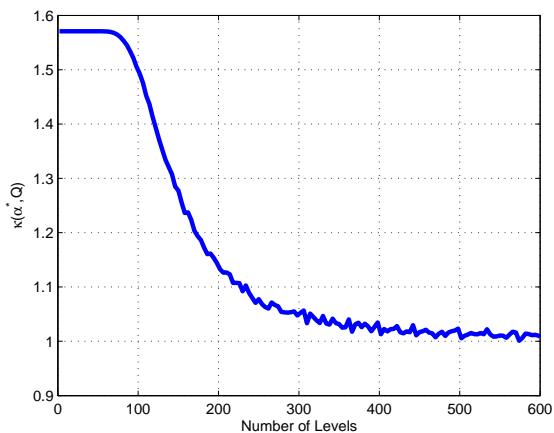} &
\widgraph{0.5\textwidth}{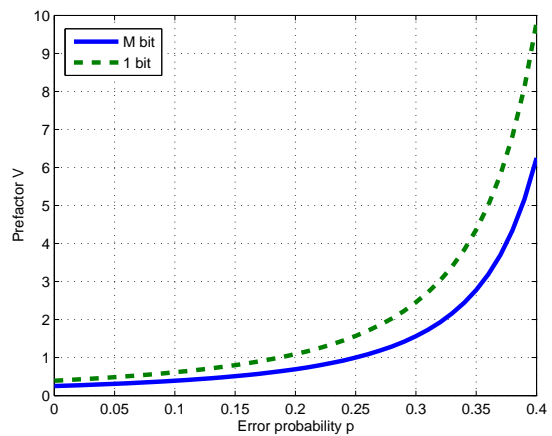} \\
(a) & (b)
\end{tabular}
\caption{(a) Plots of the asymptotic variance $\kappa(\qlevel,\Qfunc)$
defined in equation~\eqref{EqnDefnKappa} versus the number of levels
$\numlevel$ in a uniform quantizer, corresponding to $\log_2 (2
\numlevel)$ bits of feedback, for a sensor network with $\numnode =
4000$ nodes.  The plots show the asymptotic variance rescaled by the
centralized gold standard, so that it starts at $\pi/2$ for $\numlevel
= 2$, and decreases towards $1$ as $\numlevel$ is increased towards
$\numnode/2$.  (b) Plots of the asymptotic variances
$V_\numnode(\chanflip)$ and $V_1(\chanflip)$ defined in
equation~\eqref{EqnDefnVnoise} as the feedforward noise parameter
$\chanflip$ is increased from $0$ towards $\frac{1}{2}$.}
\label{FigLevels}
\end{center}
\end{figure}

\subsection{Extensions to noisy  links}

We now briefly consider the effect of communication noise on our
algorithms.  There are two types of noise to consider: (a)
\emph{feedforward}, meaning noise in the link from sensor node to
fusion center, and (b) \emph{feedback}, meaning noise in the feedback
link from fusion center to the sensor nodes.  Here we show that
feedforward noise can be handled in a relatively straightforward way
in our algorithmic framework.  On the other hand, feedback noise
requires a different analysis, as the different sensors may loose
synchronicity in their updating procedure. Although a thorough
analysis of such asynchronicity is an interesting topic for future
research, we note that assuming noiseless feedback is not
unreasonable, since the fusion center typically has greater
transmission power.

Focusing then on the case of feedforward noise, let us assume that the
link between each sensor and the fusion center acts as a binary
symmetric channel (BSC) with probability $\chanflip \in [0,
\frac{1}{2})$.  More precisely, if a bit $x \in \{0,1\}$ is
transmitted, then the received bit $y$ has the (conditional)
distribution
\begin{eqnarray}
\label{EqnDefnBSC}
\prob(y \, \mid \, x) & = & \begin{cases}
1-\chanflip & \mbox{if $x= y$} \\
\chanflip & \mbox{if $x \neq y$}.
                \end{cases}
\end{eqnarray}
With this bit-flipping noise, the updates (both
equation~\eqref{EqnMbf} and~\eqref{Eqn1bf}) need to be modified so as
to correct for the bias introduced by the channel noise.  If $\qlevel$
denotes the desired quantile, then in the presence of BSC($\chanflip)$
noise, both algorithms should be run with the modified parameter
\begin{eqnarray}
\label{EqnDefnQlevmod}
\qlevmod & \defn & (1-2 \chanflip) \qlevel + \chanflip.
\end{eqnarray}
Note that $\qlevmod$ ranges between $\qlevel$ (for the noiseless case
$\chanflip = 0$), to a quantity arbitrarily close to $\frac{1}{2}$, as
the channel approaches the extreme of pure noise ($\chanflip =
\frac{1}{2}$).  The following lemma shows that for all $\chanflip <
\frac{1}{2}$, this adjustment~\eqref{EqnDefnQlevmod} suffices to
correct the algorithm.  Moreover, it specifies how the resulting
asymptotic variance depends on the noise parameter:
\begin{proposition}
\label{PropNoise}
Suppose that each of the $\numnode$ feedforward links from sensor to
fusion center are modeled as i.i.d. BSC channels with probability
$\chanflip \in [0, \frac{1}{2})$.  Then the $\numnode$-bf or $1$-bf
algorithms, with the adjusted $\qlevmod$, are strongly consistent in
computing the $\qlevel$-quantile.  Moreover, with appropriate step
size choices, their asymptotic MSEs scale as $1/(\numnode \numobs)$
with respective pre-factors given by
\begin{subequations}
\label{EqnDefnVnoise}
\begin{eqnarray}
\label{EqnDefnVnoiseM}
V_\numnode(\chanflip) & \defn & \frac{\const^2 \; \qlevmod \,
(1-\qlevmod)}{ \big[2 \const (1-2\chanflip) p_X(\thetaopt) - 1\big]}
\\
\label{EqnDefnVnoise1}
V_1(\chanflip) & \defn & \left [\frac{\const^2 \sqrt{2 \pi \qlevmod
(1-\qlevmod )}}{8 \const (1-2 \chanflip) p_X(\thetaopt) - 4\sqrt{2 \pi
\qlevmod (1-\qlevmod )}} \right].
\end{eqnarray}
\end{subequations}
In both cases, the asymptotic MSE is minimal for $\chanflip = 0$.
\end{proposition}

\mybeginproof If sensor node $i$ transmits a bit $\transbit_{n+1}(i)$
at round $n+1$, then the fusion center receives the random variable
\begin{eqnarray*}
\widetilde{\transbit}_{n+1}(i) & = & \transbit_{n+1}(i) \oplus
W_{n+1},
\end{eqnarray*}
where $W_{n+1}$ is Bernoulli with parameter $\chanflip$, and $\oplus$
denotes addition modulo two.  Since $W_{n+1}$ is independent of the
transmitted bit (which is Bernoulli with parameter $F(\theta_n)$), the
received value $\widetilde{\transbit}_{n+1}(i)$ is also Bernoulli,
with parameter
\begin{equation}
\label{EqnBerConv}
\chanflip \ast F(\theta_n) \; = \; \chanflip \, \left (1-F(\theta_n)
\right)+ (1-\chanflip) \, F(\theta_n) \; = \; \chanflip + \left(1 - 2
\chanflip \right) F(\theta_n).
\end{equation}
Consequently, if we set $\qlevmod$ according to
equation~\eqref{EqnDefnQlevmod}, both algorithms will have their
unique fixed point when $F(\theta) = \qlevel$, so will compute the
$\qlevel$-quantile of $X$.  The claimed form of the asymptotic
variances follows from by performing calculations analogous to the
proofs of Theorems~\ref{ThmMbf} and~\ref{Thm1bf}.  In particular, the
partial derivative with respect to $\theta$ now has a multiplicative
factor $(1-2 \chanflip)$, arising from equation~\eqref{EqnBerConv} and
the chain rule.  To establish that the asymptotic variance is
minimized at $\chanflip = 0$, it suffices to note that the derivative
of the MSE with respect to $\chanflip$ is positive, so that it is an
increasing function of $\chanflip$.

\myendproof

Of course, both the algorithms will fail, as would be expected, if
$\chanflip=1/2$ corresponding to pure noise.  However, as summarized
in Proposition~\ref{PropNoise}, as long as $\chanflip < \frac{1}{2}$,
feedforward noise does not affect the asymptotic rate itself, but
rather only the pre-factor in front of the $1/(\numnode \numobs)$
rate.  Figure~\ref{FigLevels}(b) shows how the asymptotic variances
$V_\numnode(\chanflip)$ and $V_1(\chanflip)$ behave as $\chanflip$ is
increased towards $\chanflip = \frac{1}{2}$.

\section{Discussion}
\label{SecDiscussion}

 In this paper, we have proposed and analyzed different approaches to
the problem of decentralized quantile estimation under communication
constraints.  Our analysis focused on the fusion-centric architecture,
in which a set of $\nodenum$ sensor nodes each collect an observation
at each time step.  After $\numobs$ rounds of this process, the
centralized oracle would be able to estimate an arbitrary quantile
with mean-squared error of the order $\mathcal{O}(1/(\numnode
\numobs))$.  In the decentralized formulation considered here, each
sensor node is allowed to transmit only a single bit of information to
the fusion center.  We then considered a range of decentralized
algorithms, indexed by the number of feedback bits that the fusion
center is allowed to transmit back to the sensor nodes.  In the
simplest case, we showed that an $\log \numnode$-bit feedback
algorithm achieves the same asymptotic variance $\order(1/(\numnode
\numobs))$ as the centralized estimator.  More interestingly, we also
showed that that a $1$-bit feedback scheme, with suitably designed
step sizes, can also achieve the same asymptotic variance as the
centralized oracle.  We also showed that using intermediate amounts of
feedback (between $1$ and $\numnode$ bits) does not alter the scaling
behavior, but improves the constant.  Finally, we showed how our
algorithm can be adapted to the case of noise in the feedforward links
from sensor nodes to fusion center, and the resulting effect on the
asymptotic variance.

Our analysis in the current paper has focused only on the fusion
center architecture illustrated in Figure~\ref{FigNetwork}.  A natural
generalization is to consider a more general communication network,
specified by an undirected graph on the sensor nodes.  One possible
formulation is to allow only pairs of sensor nodes connected by an
edge in this communication graph to exchange a bit of information at
each round.  In this framework, the problem considered in this paper
effectively corresponds to the complete graph, in which every node
communicates with every other node at each round.  This more general
formulation raises interesting questions as to the effect of graph
topology on the achievable rates and asymptotic variances.

\subsection*{Acknowledgements}
We would like to thank Prof. Pravin Varaiya for some discussion that
led to the initial ideas on this subject. RR was supported by the
California Department of Transportation through the California PATH
program. MJW was partially supported by NSF grant DMS-0605165 and an
NSF CAREER award CCF-0545862.


\appendix

\noindent {\bf{\large{Proof of Theorem~\ref{ThmGen}:}}}

\label{AppGen}

We proceed in an analogous manner to the proof of
Theorem~\ref{ThmMbf}:
\begin{lemma} \label{fQ_mon_dec}
For fixed $x \in [0,1]$, the function $\EQfunc(r,x)$ is non-negative,
differentiable and monotonically decreasing.
\end{lemma}

\mybeginproof
First notice that by definition:
\begin{equation}
\EQfunc(r,x) = \expec\Big[\Qfunc \left[x-\frac{X}{\numnode} \right]
\Big ],
\end{equation}
where $X$ is a $Bin(r,\numnode)$ random variable. Note that if $X'
\sim Bin(r',\numnode)$, with $r'>r$, then certainly $\prob\left(X'\leq
n\right) \leq \prob\left(X \leq n\right)$---meaning that $X'$
stochastically dominates $X$. For any constant $x$,
$\prob\left(x-\frac{X'}{\numnode}\leq s\right) \geq
\prob\left(x-\frac{X}{\numnode} \leq s\right)$. Furthermore, by the
quantizer is, by definition, a monotonically non-decreasing function.
Consequently, a standard result on stochastic domination~\cite[\S
4.12]{Grimmett} implies that $\EQfunc(r,x) \geq
\EQfunc(r',x)$. Differentiability follows from the definition of the
function.

\myendproof

The finiteness of the variance of the quantization step is clear by
construction; more specifically, a crude upper bound is
$r_\numlevel^2$. Thus, analogous to the previous theorems,
Lemma~\ref{fQ_mon_dec} is used to establish almost sure convergence.

Now, some straightforward algebra using the results of
Lemma~\ref{f_mon_rate} shows that the partial derivative
$\frac{\partial \EQfunc(r,x)}{\partial r}$ is
\begin{equation}
\label{pdfull} \frac{1}{r(1-r)} \sum_{k = -\numlevel}^{\numlevel-1}
r_k \: \left\{ \expec \left [X \; \ind \left( x-s_{k+1} \leq
\frac{X}{\nodenum} \leq x-s_{k} \right) \right] - \expec[X] \; \Prob
\left[x-s_{k+1} \leq \frac{X}{\nodenum} \leq x-s_{k} \right] \right
\},
\end{equation}

This will be used next. To compute the asymptotic variance, we again
exploit asymptotic normality (see equation~\eqref{EqnNormAs}) as
before:
\begin{eqnarray*}
\expec[X \ind(\nodenum(\qlevel-s_{k+1})\leq X\leq \nodenum
(\qlevel-s_{k}))] & = & \expec\left[X \ind\left(-\sqrt{m}s_{k+1}\leq
\frac{X-\qlevel\nodenum} {\sqrt{m}} \leq
-\sqrt{\nodenum}s_{k}\right)\right]\nonumber\\
& = &
\sqrt{m}\expec\left[(Z+\qlevel\sqrt{\nodenum})\ind\left(-\sqrt{m}s_{k+1}\leq
Z \leq -\sqrt{\nodenum}s_{k}\right)\right]\nonumber\\
& = &\sqrt{m}\expec\left[Z \ind\left(-\sqrt{m}s_{k+1}\leq Z \leq
  -\sqrt{\nodenum}s_{k}\right)\right]+S\nonumber\\
& \rightarrow & -\sqrt{\nodenum}\int_{\sqrt{m}s_{k}}^{\sqrt{m}s_{k+1}}
z \frac{\exp\left(\frac{-z^2}{2a}\right)}{\sqrt{2\pi a}} dz+S
\nonumber\\
S &\defn& \expec[X]P(\nodenum(x-s_{k+1})\leq X\leq \nodenum(x-s_{k}))
\end{eqnarray*}

Now make the definition, which corresponds to solving the integral
above:

\begin{eqnarray*}
\Delta_{\numnode}(s_{k},s_{k+1}) = \left(\exp\left(-\frac{\numnode
s_{k}^2}{2\qlevel(1-\qlevel)}\right)-\exp\left(-\frac{\numnode
s_{k+1}^2}{2\qlevel(1-\qlevel)}\right)\right)
\end{eqnarray*}

Thus, plugging into Equation~\ref{pdfull}, noticing that $S$
cancels:

\begin{eqnarray*}
\frac{\partial \EQfunc(r,\qlevel)}{\partial r} \big|_{r =
F(\thetaopt)} \rightarrow  -\sqrt{\frac{\nodenum}{2\pi \qlevel
(1-\qlevel)}}\sum_{k =
-\numlevel}^{\numlevel-1}r_k\Delta_{\numnode}(s_{k},s_{k+1})
\end{eqnarray*}

A side note is that if one chooses $s_0 =0$, we are guaranteed that
at least one $\Delta_{\numnode}(s_{k},s_{k+1})$ does not go to zero
in a fixed quantizer (i.e. a quantizer where the levels $s_k$ do not
depend on $\numnode$). But the correction factor expression, and as
a matter of fact, the optimum quantization of Gaussian, suggests
that the levels $s_{k}$ scale as $1/\sqrt{\numnode}$. In this case,
the factor is a constant, independent of $\numnode$.

We now need to compute $R(\thetaopt)$ for the quantized updated. It
is also straightforward to see that this quantity is given by:

\begin{eqnarray*}
R(\thetaopt) = \const_{\numnode}^2\sum_{k =
-\numlevel}^{\numlevel-1}r_k^2 (\Gfunc(F(\thetaopt),
\qlevel-s_{k})-\Gfunc(F(\thetaopt),
\qlevel-s_{k+1}))-\specpar^2\nonumber
\end{eqnarray*}

Putting everything together we obtain the asymptotic variance
estimate for the more general quantizer converges to:

\begin{eqnarray*}
\frac{R(\thetaopt)}{2 \const_m \left |\frac{\partial \EQfunc(r,
\thetaopt)}{\partial r} \big |_{r = \qlevel} \right| p_X(\thetaopt)
- 1} \rightarrow\nonumber\\
\frac{\const_{\numnode}^2\sum_{k = -\numlevel}^{\numlevel-1}r_k^2
(\Gfunc(F(\thetaopt), \qlevel-s_{k})-\Gfunc(F(\thetaopt),
\qlevel-s_{k+1}))-\specpar^2}{\frac{2\const_{\numnode}\sqrt{\numnode}\sum_{k
=-\numlevel}^{\numlevel-1}r_k\Delta_{\numnode}(s_{k},s_{k+1})p_X(\thetaopt)}{\sqrt{2\pi
\qlevel (1-\qlevel)}}-1}
\end{eqnarray*}

Set a gain $K = \frac{\const_{\numnode}\sqrt{\numnode}\sum_{k
=-\numlevel}^{\numlevel-1}r_k\Delta_{\numnode}(s_{k},s_{k+1})}{\sqrt{2\pi
\qlevel (1-\qlevel)}}$ and we have the final expression for the
variance:

\begin{eqnarray*}
2\pi\frac{\sum_{k =
-\numlevel}^{\numlevel-1}r_k^2\Delta\Gfunc(s_{k},s_{k+1})
-\specpar^2}{\left(\sum_{k
=-\numlevel}^{\numlevel-1}r_k\Delta_{\numnode}(s_{k},s_{k+1})\right)^2}\left[\frac{\const^2
\qlevel(1-\qlevel)}{2 K p_X(\thetaopt)-1}\frac{1}{\numnode}\right]
\end{eqnarray*}

Where $\Delta\Gfunc(s_{k},s_{k+1}) = \Gfunc(\qlevel,
\qlevel-s_{k})-\Gfunc(\qlevel, \qlevel-s_{k+1})$. The constant
$\kappa(\qlevel,\Qfunc)$ defines the performance of the algorithm
for different quantization choices:

\begin{eqnarray*}
\kappa(\qlevel,\Qfunc) = 2\pi\frac{\sum_{k =
-\numlevel}^{\numlevel-1}r_k^2\Delta\Gfunc(s_{k},s_{k+1})
-\specpar^2}{\left(\sum_{k
=-\numlevel}^{\numlevel-1}r_k\Delta_{\numnode}(s_{k},s_{k+1})\right)^2}
\end{eqnarray*}

The rate with respect to $\numnode$ is the same, independent of
quantization. It is clear from previous analysis that if the best
quantizers are chosen $1 \leq \kappa(\qlevel,\Qfunc) \leq
\frac{2\pi}{4}$. Obviously $\kappa(\qlevel,\Qfunc)$ over the class
of optimal quantizers is a decreasing function of $\numlevel$.

\bibliographystyle{plain}

\begin{thebibliography}{10}

\bibitem{Amari89}
S.~Amari and T.~S. Han.
\newblock Statistical inference under multiterminal rate restrictions: A
  differential geometric approach.
\newblock {\em IEEE Trans. Info. Theory}, 35(2):217--227, March 1989.

\bibitem{Ayanoglu90}
E.~Ayanoglu.
\newblock On optimal quantization of noisy sources.
\newblock {\em IEEE Trans. Info. Theory}, 36(6):1450--1452, 1990.

\bibitem{Benveniste90}
A.~Benveniste, M.~Metivier, and P.~Priouret.
\newblock {\em Adaptive Algorithms and Stochastic Approximations}.
\newblock Springer-Verlag, New York, NY, 1990.

\bibitem{Blum97}
R.~S. Blum, S.~A. Kassam, and H.~V. Poor.
\newblock Distributed detection with multiple sensors: Part ii---advanced
  topics.
\newblock {\em Proceedings of the IEEE}, 85:64--79, January 1997.

\bibitem{Chamberland04}
J.~F. Chamberland and V.~V. Veeravalli.
\newblock Asymptotic results for decentralized detection in power constrained
  wireless sensor networks.
\newblock {\em IEEE Journal on Selected Areas in Communication},
  22(6):1007--1015, August 2004.

\bibitem{Chong03}
C.~Chong and S.~P. Kumar.
\newblock Sensor networks: Evolution, opportunities, and challenges.
\newblock {\em Proceedings of the IEEE}, 91:1247--1256, 2003.

\bibitem{Grimmett}
G.R. Grimmett and D.R. Stirzaker.
\newblock {\em Probability and random processes}.
\newblock Oxford Science Publications, Clarendon Press, Oxford, 1992.

\bibitem{Gubner93}
J.~A. Gubner.
\newblock Decentralized estimation and quantization.
\newblock {\em IEEE Trans. Info. Theory}, 39(4):1456--1459, 1993.

\bibitem{Han90}
J.~Han, P.~K. Varshney, and V.~C. Vannicola.
\newblock Some results on distributed nonparametric detection.
\newblock In {\em Proc. 29th Conf. on Decision and Control}, pages 2698--2703,
  1990.

\bibitem{Han98}
T.~S. Han and S.~Amari.
\newblock Statistical inference under multiterminal data compression.
\newblock {\em IEEE Trans. Info. Theory}, 44(6):2300--2324, October 1998.

\bibitem{Han89}
T.~S. Han and K.~Kobayashi.
\newblock Exponential-type error probabilities for multiterminal hypothesis
  testing.
\newblock {\em IEEE Trans. Info. Theory}, 35(1):2--14, January 1989.

\bibitem{Kushner97}
H.~J. Kushner and G.~G. Yin.
\newblock {\em Stochastic Approximation Algorithms and Applications}.
\newblock Springer-Verlag, New York, NY, 1997.

\bibitem{Luo05}
Z.Q. Luo.
\newblock Universal decentralized estimation in a bandwidth-constrained sensor
  network.
\newblock {\em IEEE Trans. Info. Theory}, 51(6):2210--2219, 2005.

\bibitem{NguWaiJor05}
X.~Nguyen, M.~J. Wainwright, and M.~I. Jordan.
\newblock Nonparametric decentralized detection using kernel methods.
\newblock {\em IEEE Trans. Signal Processing}, 53(11):4053--4066, November
  2005.

\bibitem{Serfling80}
R.~J. Serfling.
\newblock {\em Approximation Theorems of Mathematical Statistics}.
\newblock Wiley Series in Probability and Statistics. Wiley, 1980.

\bibitem{Tenney81}
R.~R. Tenney and N.~R.~Jr. Sandell.
\newblock Detection with distributed sensors.
\newblock {\em IEEE Trans. Aero. Electron. Sys.}, 17:501--510, 1981.

\bibitem{Tsitsiklis93}
J.~N. Tsitsiklis.
\newblock Decentralized detection.
\newblock In {\em Advances in Statistical Signal Processing}, pages 297--344.
  JAI Press, 1993.

\bibitem{Veeravalli93}
V.~V. Veeravalli, T.~Basar, and H.~V. Poor.
\newblock Decentralized sequential detection with a fusion center performing
  the sequential test.
\newblock {\em IEEE Trans. Info. Theory}, 39(2):433--442, 1993.

\bibitem{Viswanathan97}
R.~Viswanathan and P.~K. Varshney.
\newblock Distributed detection with multiple sensors: Part i---fundamentals.
\newblock {\em Proceedings of the IEEE}, 85:54--63, January 1997.

\bibitem{Zhang88}
Z.~Zhang and T.~Berger.
\newblock Estimation via compressed information.
\newblock {\em IEEE Trans. Info. Theory}, 34(2):198--211, 1988.

\bibitem{Zie04}
R.~Zielinski.
\newblock Optimal quantile estimators: Small sample approach.
\newblock Technical report, Inst. of Math. Pol. Academy of Sci., 2004.

\end{thebibliography}

\end{document}